\documentclass[RNAAS]{aastex62}
\usepackage{amsmath, amssymb, bm}

\begin{document}

\title{A catalog of 29 open clusters and associations observed by the {\em Kepler} and {\em K2} Missions}

\correspondingauthor{Ann Marie Cody}
\email{annmarie.cody@nasa.gov}

\author{Ann Marie Cody}
\affiliation{NASA Ames Research Center, Moffett Blvd, Mountain View, CA 94035, USA}
\affiliation{Bay Area Environmental Research Institute, 625 2nd St Ste. 209, Petaluma, CA 94952}

\author{Geert Barentsen}
\affiliation{NASA Ames Research Center, Moffett Blvd, Mountain View, CA 94035, USA}
\affiliation{Bay Area Environmental Research Institute, 625 2nd St Ste. 209, Petaluma, CA 94952}

\author{Christina Hedges}
\affiliation{NASA Ames Research Center, Moffett Blvd, Mountain View, CA 94035, USA}
\affiliation{Bay Area Environmental Research Institute, 625 2nd St Ste. 209, Petaluma, CA 94952}

\author{Michael Gully-Santiago}
\affiliation{NASA Ames Research Center, Moffett Blvd, Mountain View, CA 94035, USA}
\affiliation{Bay Area Environmental Research Institute, 625 2nd St Ste. 209, Petaluma, CA 94952}

\author{Jessie Dotson}
\affiliation{NASA Ames Research Center, Moffett Blvd, Mountain View, CA 94035, USA}
\affiliation{Bay Area Environmental Research Institute, 625 2nd St Ste. 209, Petaluma, CA 94952}

\author{Thomas Barclay}
\affiliation{NASA Goddard Space Flight Center, 8800 Greenbelt Rd., Greenbelt, MD 20771, USA}

\author{Steve Bryson}
\affiliation{NASA Ames Research Center, Moffett Blvd, Mountain View, CA 94035, USA}

\author{Nicholas Saunders}
\affiliation{NASA Ames Research Center, Moffett Blvd, Mountain View, CA 94035, USA}
\affiliation{Bay Area Environmental Research Institute, 625 2nd St Ste. 209, Petaluma, CA 94952}


\section*{Kepler's Cluster Observations}


Since its launch in 2009, the {\em Kepler Space Telescope} (Borucki et al.\ 2010) has photometrically monitored $\sim$500,000 stars comprising a 
large range of ages. Its 4-year ``prime'' mission observed a single $\sim 10\arcdeg\times 10$\arcdeg\ field in Cygnus, while the {\em K2} Mission 
(initiated in 2014; Howell et al.\ 2014) has focused for 70--80 days at a time on objects in the ecliptic plane. A highlight of {\em Kepler's} 
observations has been the inclusion of galactic open and globular clusters in its target list. These groups of stars are considered to be coeval, 
and thus provide ideal laboratories for testing stellar astrophysics and planet occurrence versus age.

The {\em Kepler} and {\em K2} Missions publicly release all collected imaging data, as well as light curves for the majority of targets. Cluster observations are collected in one of two formats: 1.) small ``target pixel files'' (TPFs)  centered on isolated single targets, or 2.) ``superstamps''-- larger regions consisting of both the cluster of interest and foreground/background stars (see Cody et al.\ 2018). The missions have produced and released light curves for targets in TPFs, but have left it up to the community to extract them from superstamps. Stars in crowded cluster environments require more careful tuning of photometric analysis methods to produce precision time series. As a result, some {\em Kepler/K2} clusters have received more attention than others to date. In this note, we present the compilation of all clusters and associations observed by {\em Kepler} from 2009--2018, and highlight key cluster science results as well as potential avenues for future research.

\begin{deluxetable}{llcccc}
\tablecolumns{6}
\tablewidth{0pt}
\tablecaption{Clusters and associations observed during the {\em Kepler} and {\em K2} Missions}
\tablehead{
\colhead{Region} & \colhead{Campaign} & \colhead{Super-} & \colhead{Region} & \colhead{Age Range} & \colhead{Extensively}\\
\colhead{} & \colhead{} & \colhead{stamp?} & \colhead{Type} & \colhead{[Gyr]} & \colhead{studied?}\\
}
\startdata
  NGC 6819 & Kepler prime& Y & open cluster& 1--2 & Y\\
  NGC 6791 & Kepler prime& Y & open cluster& $\sim$8 & Y\\
  NGC 6811 & Kepler prime& N & open cluster& 1--2 & N\\
  NGC 6866 & Kepler prime& N & open cluster& 0.4--0.8 & N\\
  M35 & K2 C0 & Y & open cluster& 0.1--0.2 & N\\
  NGC 2158 & K2 C0 & Y & open cluster& 1--2 & N\\
  M4 & K2 C2& Y & globular cluster& $\gtrsim$11 & N\\
  M80 & K2 C2& Y & globular cluster& $\gtrsim$11 & N\\
  Upper Scorpius & K2 C2, C15& N & association & $\lesssim$0.01 & Y\\
  $\rho$ Ophiuchus & K2 C2& N & open cluster & $<$0.01 & N\\
  Pleiades & K2 C4 & N & open cluster & 0.1--0.2 & Y\\
  Hyades & K2 C4, C13& N & open cluster & 0.4--0.8 & Y\\
  M67 & K2 C5, C16, C18& Y & open cluster & 3--4 & Y\\
  Praesepe & K2 C5, C16, C18& N & open cluster & 0.4--0.8 & Y\\
  Ruprecht 147 & K2 C7 & Y & open cluster & 3--4 & N\\
  NGC 6717 & K2 C7& N & globular cluster & $\gtrsim$11 & N\\
  NGC 6530 & K2 C9& Y & open cluster & $<$0.01 & N\\
  M9 & K2 C11& Y & globular cluster & $\gtrsim$11 & N\\
  M19 & K2 C11& Y & globular cluster & $\gtrsim$11 & N\\
  NGC 6293 & K2 C11& Y & globular cluster & $\gtrsim$11 & N\\
  NGC 6355 & K2 C11& Y & globular cluster & $\gtrsim$11 & N\\
  Terzan 5 & K2 C11& Y & globular cluster & $\gtrsim$11 & N\\
  NGC 1647 & K2 C13& N & open cluster & 0.1--0.2 & N\\
  NGC 1746 & K2 C13& N & open cluster* & - & N\\
  NGC 1817 & K2 C13& N & open cluster & 1--2 & N\\
  Taurus & K2 C13& N & association & $<$0.01 & N\\
  NGC 1750 & K2 C13& N & open cluster & 0.1--0.2 & N\\
  NGC 1758 & K2 C13& N & open cluster & 0.4--0.8 & N\\
  NGC 5897 & K2 C15& Y & globular cluster & $\gtrsim$11 & N\\
\enddata
\tablecomments{Basic data on the clusters and associations photometrically monitored as part of the {\em Kepler} and {\em K2} missions. Column six provides a measure of how well studied the cluster is by noting whether there are more than five papers analyzing its {\em Kepler} or {\em K2} data. We note that the cluster status of NGC 1746 is debated; it may be an asterism of unrelated stars.}
\end{deluxetable}

\section*{Cluster Science and Future Potential}

In all, {\em Kepler} has observed two star forming associations (age $<$10~Myr), 18 open clusters (1~Myr to 8~Gyr), and nine globular clusters ($\gtrsim$11~Gyr). The compilation of clusters is provided in Table 1, along with the observing campaign and an assessment of the degree of attention received by each cluster's {\em Kepler/K2} data in the literature. 

Investigations into {\em Kepler/K2} clusters have comprised a number of areas, from stellar to planetary science. Asteroseismology, particularly focusing on red giants, has been carried out in the open clusters observed during the {\em Kepler} prime mission (Hekker et al.\ 2011, Corsaro et al.\ 2012, and references therein), and more recently, the first solar type oscillations have been detected in a {\em K2} cluster (the Hyades; Lund et al.\ 2016). Asteroseismology has also been conducted in M67 (Stello et al.\ 2016) and M4, one of the globular clusters targeted by {\em K2} (Miglio et al.\ 2016).

Many cluster members have starspots, making gyrochronology and rotation studies popular endeavors. Stellar spin-down has been probed in the main sequence stars of M67 (Barnes et al. 2016 and references therein) as well as the 2.5 billion-year-old cluster NGC~6819 (Meibom et al. 2015). Stellar angular momentum evolution from the pre-main sequence to intermediate ages has also received attention, with analyses by Rebull et al.\ (2016, 2018) among others. 

Among the clusters, numerous eclipsing binary systems have been uncovered and used to probe stellar structure as a function of age (e.g., Gillen et al.\ 2017 and references therein). The {\em Kepler} mission also enabled the first discovery of planets in an open cluster (Meibom et al.\ 2013), leading to a handful of other finds from {\em K2} (Livingston et al.\ 2018, Vanderburg et al.\ 2018 and references therein). In the youngest clusters and associations, precision time series photometry has also enabled detailed analyses of pre-main sequence accretion phenomena and obscuration by circumstellar material (Cody \& Hillenbrand 2018 and references therein).

While the science return thus far from the clusters of {\em Kepler} and {\em K2} has been excellent, only a fraction of its potential has been mined. According to our criteria of assessment ($>$5 refereed papers), only a third of observed clusters have been extensively studied thus far. We hope that by providing the complete list observed by {\em Kepler}, the astronomical community will be motivated to further probe the astrophysics of member stars and their companions. The more crowded regions will benefit from new photometry techniques under development, such as psf modeling and fitting. This is especially the case for the globular clusters, which provide an opportunity to study some of the oldest stars in the universe.

\end{document}